# Electrically Conductive GNP/Epoxy Composites for Out-of-Autoclave Thermoset Curing through Joule Heating

Tian Xia, Desen Zeng, Zheling Li, Robert J. Young, Cristina Vallés* and Ian A. Kinloch*

School of Materials and National Graphene Institute, University of Manchester, Oxford Road, Manchester, M13 9PL, UK

*cristina.valles@manchester.ac.uk; ian.kinloch@manchester.ac.uk

**Abstract**

The development of scalable Out-of-Autoclave (OoA) *in-situ* thermoset curing methods are required to overcome important drawbacks related to the autoclave-based processing methods typically used in industry. The incorporation of graphene, an electrothermal carbon nanomaterial with the ability to transform electric energy into heat through Joule heating, emerges as a promising route to replace the conventional processing methods. In this work the electrical behaviour of both uncured and oven cured GNPs/epoxy composites with loadings of up to 10 wt.% were evaluated and electrical percolation thresholds were established for both. Above the critical loading found for oven cured materials (~ 8.5 wt.%) the electrically conducting networks of GNPs formed in the matrix showed the ability to act as integrated nanoheaters when an electric current was passed through them, successfully curing the composites by Joule heating. Composites prepared by this OoA curing method (as an alternative to the traditional oven based one) at 10 wt.% loading of GNPs were also prepared and compared to the oven cured ones. They showed more compact composite structures, with less microvoids and a preferred orientation of the GNPs in the matrix relative to the oven cured material at identical loading, as revealed by electron microscopy and polarized Raman spectroscopy, respectively. This microstructure and anisotropy induced by the electrically-induced (i.e. OoA) cure led to GNPs/epoxy composites with superior electrical and mechanical properties (revealed by tensile testing). The well-distributed GNP nanoparticles acting as nanoheaters integrated in a thermosetting matrix, in combination with excellent mechanical and electrical performances achieved for the overall graphene/epoxy composites and the simplicity associated to the method, should open the door to novel industrial applications.

**1. Introduction**

There is an increasing demand for high-performance polymer matrix composite materials in many industrial sectors such as defense, marine, aerospace and automotive. In these sectors thermosetting polymers are mostly used, that require thermal energy to activate their polymerization. Thus a conventional oven or autoclave is typically employed to cure the thermosetting composite during manufacture. Whilst these manufacturing techniques are effective for high-value, low to medium throughput applications, there are important drawbacks associated to them, including geometry limitation, massive energy consumption, inflexible operation conditions and inability to repair in the field. Thus there is a strong motivation to develop scalable Out-of-Autoclave (OoA) curing methods [1].

One promising OoA route is the incorporation of electrothermal materials that transform electric energy into heat through Joule heating of the composite. Joseph and Viney showed that carbon fibre/epoxy composites could be successfully cured by the ohmic heating generated by passing an electrical current through the fibres [2]. They reported that the properties of resistance cured composite compared favourably with those of oven-cured samples, and that significantly less energy was required in the process of resistance curing small samples, relative to curing in an oven or autoclave. Nanomaterials, and particularly carbon-based ones, that offer more and much better distributed hot spots in the matrix, emerge as highly advantageous and promising materials for the development of OoA cure methods based on the use of nanocomposites. For example, recently, a self-heating fiber reinforced polymer composite using meso/macropore carbon nanotube (CNT) paper was sucessfully fabricated and its application in deicing was reported [3].

Among all carbon nanomaterials, graphene seems particularly promising for this application. Important improvements on the mechanical and electrical properties of epoxy matrices by incorporating different types of graphene have already been reported for the fabrication of structural composites for aerospace applications [4-7]. In addition, above the electrical percolation threshold the well-distributed high-aspect ratio graphene nanoparticles form electrically conductive networks in the matrix with great potential as integrated nanostructured resistive heaters. Mas *et al.* [8] have recently developed a thermoset curing method through Joule heating of nanocarbons (which includes graphene) for composite manufacture, repair and soldering. Prolongo et al. [9] have also recently reported the Joule effect self-heating of epoxy composites reinforced with graphene nanoplatelets (GNPs) and CNTs, which successfully led to the cure of thermoset systems.

Although CNTs typically form percolated networks at lower loadings than GNPs, due to their higher surface areas and higher thermal conductivities relative to CNTs [10, 11], GNPs emerge as particularly interesting as carbon nanofiller to develop an electrically induced method to cure in-situ thermoset systems. In addition, graphene materials are easier to process in terms of viscosities (lower viscosities are typically achieved with GNPs relative to CNTs at identical loadings) and they offer enhanced adhesion with the host polymer relative to CNTs. They are also typically associated to additional functionality that CNTs or other types of nanocarbon fillers cannot offer, such as barrier properties or exceptional electromagnetic interference (EMI) shielding performances. GNPs/epoxy composites hold thus great promise for novel applications for which additional functionality might be required besides exceptional mechanical and electrical properties.

An investigation of the electrical behavior of GNPs/epoxy uncured and cured composites and a detailed comparison between the microstructure and properties of the composites fabricated using the conventional oven and the Joule heating curing method, as well as a detailed study of how the electrical conductance develops in graphene/epoxy composites during curing has not yet, however, been performed. We consider this of great importance to develop and control in-situ electrically induced curing protocols for thermosetting systems.

In this work the electrical behaviour of both uncured and oven cured GNPs/epoxy composites were evaluated and percolation thresholds were established for the first time.

Above the electrical percolation threshold found for the oven cured materials, the efficiency of the electrically conductive well-distributed GNP nanoparticles acting as nanoheaters integrated in a thermoset matrix to *in-situ* cure thermoset systems by simple Joule heating when an electric current is passed through them was evaluated. How this novel OoA method compares with the traditional oven-based one in terms of microstructure, electrical and mechanical properties of the GNP/epoxy composites at identical loading was investigated. Besides, in contrast to previous works, which are not evaluating or giving any evidence of the effect of the applied electric field on the alignment of the fillers in the epoxy matrix, we are introducing here an evaluation of the alignment induced by the electric current passing through the material during curing, which represents an important novelty with respect to previous works reporting similar OoA curing methods. The potential that this OoA method has to be scaled-up industrially is also discussed.

## 2. Experimental

*2.1. Preparation of the GNPs/epoxy composites.*

Graphene Nanoplatelets (GNP) M25 from XG Sciences, with lateral dimension and thickness quoted by the manufacturer of ~ 25 µm and ~ 6 nm, respectively, was used in this work as nanofiller. The epoxy resin (Araldite, LY5052) and the hardener (Aradur, HY5052) were purchased from Huntsman. Initially, the percolation threshold was found for both uncured GNP/epoxy mixtures and oven-cured GNP/epoxy composites (the later denoted as 'GNP/epoxy-Oven'). The electrical conductivity was also evaluated at different stages of curing. Oven-cured composites with loadings from 0.3 to 10 wt.% were prepared using a solvent dispersion method, in which the GNP powder was dispersed in acetone (0.5 mg/mL) using a sonication bath for 2.5 h. The epoxy resin was then added to the dispersion in the appropriate ratio and the mixture was sonicated for further 2.5 h. The mixture was left under mechanical stirring at room temperature overnight to remove the acetone. The hardener was then added to the mixture (38 wt.% with respect to the resin), mixed manually for several minutes and degassed under vacuum for 20 min. (Due to increases in viscosities, for loadings above 4 wt.% solvent-assisted sonication and shear mixing were combined to achieve homogeneous GNP/epoxy mixtures). After an initial 24 hour curing step at room temperature, the samples were then post-cured at 100 °C for 4 hours in a conventional oven. This curing protocol was chosen because previous works performed in the group [5, 12] using the same epoxy system probed to render the optimal properties of epoxy and graphene/epoxy composites. Having identified the percolation threshold at 8.5 wt.%, a 10 wt.% composite was prepared using an out-of-autoclave (OoA) method (denoted as 'GNP/epoxy-OoA'). The graphene was dispersed as before and left to cure for 24 hours at room temperature. However, instead of an oven, thermal activation energy was given by passing an electric current (maximum power ~40 V), carefully controlling it to keep the temperature constant at 100 ºC for 4 h. A FLIR T460 Infrared camera was used to measure the temperature of the composite during the application of the electric field. Further experimental details on the two curing methods are given in the SI.

*2.2. Characterisation of the microstructure of GNPs/epoxy nanocomposites.*

The fracture surface of the epoxy and the composites were characterized by scanning electron microscopy (SEM) using a Zeiss EVO60 VPSEM (15 kV acceleration voltage). Differential Scanning Calorimetry (DSC) was performed using a DSC Q100 analyzer (TA instruments) heating from room temperature to 200 °C at a rate of 10 °C/min. Three samples were tested for each experimental conditions. Raman spectroscopy of the GNPs, as well as Raman mapping of the GNPs/epoxy composites were performed using a low-power 633 nm HeNe laser in a Renishaw 2000 Raman spectrometer. For the Raman mapping analysis 121 Raman spectra were obtained over 20 × 20 μm areas of composite. The orientation of the graphene flakes in the polymer matrix was evaluated by polarized Raman spectroscopy using a backscattering geometry and a VV (vertical/vertical) combination of incident and scattered polarization, in which the directions of the incident and scattered polarization were the same. A rotational stage on the spectrometer optical microscope was used to rotate the specimen with respect to its axes. The change in the intensity of the G band was recorded as a function of the rotation angle [13].

*2.3. Characterization of the electrical and mechanical properties of the composites*

To measure the impedance of the uncured GNPs/epoxy composites the liquid mixtures were transferred to an electric pool (details of the measurement are given in the SI), whereas the impedance of cured epoxy and composites were tested on 13 mm x 13 mm x 1 mm solid specimens. All impedances were tested using a PSM 1735 Frequency Response Analyzer from Newtons4th Ltd connected with Impedance Analysis Interface (IAI) at the range of frequencies from 1 to $10^6$ Hz. The specific conductivities ($\sigma$) of the materials were calculated from the measured impedances using:

$$\sigma(\omega) = |Y^*(\omega)| \frac{t}{A} = \frac{1}{Z^*} \times \frac{t}{A} \qquad \text{Equation (1)}$$

where $Y^*(\omega)$ is the complex admittance, $Z^*$ is the complex impedance, $t$ and $A$ are the thickness and cross section area of the sample, respectively.

Stress-strain curves were obtained from dumbbell specimens (ASTM D 638 international standard) using an Instron 3365 machine equipped with an extensometer. A testing rate of 0.5 mm/min and a load cell of 5 kN were employed. 5 samples were tested for each sample.

## 3. Results and Discussion

*3.1. Electrical properties of oven-cured GNP/epoxy composites*

In order to develop novel resistive heating methods to cure GNP/epoxy composites an initial understanding of the electrical properties of the system was required. Figs.1a,b show log-log plots of the specific conductivities with increasing frequency for the uncured and oven cured composites, respectively, at different loadings of GNPs up to 10 wt.%. Very low conductivities were found for the epoxy and the uncured mixtures at GNPs loadings of 2 wt.% and 4 wt.%, whereas it was found to increase by an order of magnitude at 6 wt.% loading. This suggested that the formation of conductive pathways (percolation threshold) occurred between 4 and 6 wt.% loading of GNPs for the uncured GNP/epoxy mixtures. Above the percolation threshold the conductivities increased with loading, reaching values of 0.0026 S/m and 0.005 S/m for 9 wt.% and 10 wt.% content of GNPs, respectively. A higher percolation threshold (~ 8.5 wt.%) was found for the oven cured composites, above which the conductivities increased by 3 orders of magnitude from 8.5 wt.% ($8 \times 10^{-7}$ S/m) to 10 wt.% ($2.5 \times 10^{-4}$ S/m) loadings. Both uncured and cured materials showed the typical behaviour of two-phase (resistive and capacitive) systems: below percolation the conductivity was linearly dependent of the frequency (dielectric behavior), whereas above it the conductivity showed a plateau up to a critical frequency, which indicated the formation of conductive paths in the matrix. The curves of conductivities versus frequency (Fig. 1) showed the dominance of the resistive component at low frequencies (with conductivities independent on frequency) and the dominance of the capacitive component (with conductivities depending on frequency) at high frequencies (more details in the SI). Consequently, the conductivities of the epoxy and all oven-cured composites converged at very high frequencies. This convergence was not observed for the uncured system at loadings > 8 wt.% though, which we attribute to massive increases in the viscosity of the solution at high loadings, inducing aggregation and reducing fluid mobility.

The plot of the conductivities of the uncured and oven-cured composites at a frequency of 1 Hz as a function of the filler content (shown in Figs.1c,d) clearly revealed that both were percolated systems and, consequently, the percolation thresholds ($P_c$) could be calculated based on the percolation theory [38]. According to this theory, we can assume that the fillers are randomly distributed in the polymer matrix and the following equation can be used to determine the critical GNP loading (*i.e.* percolation threshold, $P_c$):

$$\sigma = \sigma_0 \, (P - P_c)^t \qquad \qquad \text{Equation (2)}$$

where $P$ is filler loading, $P_c$ is the percolation threshold, $t$ is conductivity exponent and $\sigma$ is the electrical conductivity.

From the data shown in Figs.1c,d values of $P_c$ = 4.5 wt.% and $P_c$ = 8.125 wt.% could be assumed to be used in Equation (2) for uncured and oven cured composites, respectively. Log-log plots of conductivity as a function of $P$-$P_c$ for both systems were then plotted (shown as inserts in Figs. 1c,d) and the corresponding linear fittings revealed a conductivity exponent of $t = 4.1$ with $R^2 = 0.83$ for the uncured GNPs/epoxy mixtures and an exponent of $t = 3.75$ with $R^2 = 0.93$ for the oven cured materials. Since the critical exponent, $t$, is directly related to the dimensionality on percolated systems, these values suggested the formation of 3D networks of GNPs in the epoxy matrix for both uncured and cured systems.

Higher conductivities and lower $P_c$ were found for the uncured system relative to the oven cured composites for identical loadings. In order to understand how the conductivity evolved during curing, the electrical properties of the composites were followed throughout the manufacturing process. The conductivity of a 10 wt.%-GNP/epoxy mixture at different stages of curing were determined and the results are shown in Fig. 2. The uncured mixture was found to be highly conductive (~ $4.9 \times 10^{-3}$ S/m), whereas after adding the hardener and performing a first curing at room temperature for 24 h the conductivity decreased two orders of magnitude (~ $2.83 \times 10^{-5}$ S/m). We related this to a reduction of the mobility of the GNP flakes during the liquid-solid transition and subsequent increase in viscosity, which must cause a reduction on the number of conductive pathways of graphene formed in the solid cured sample in comparison to the liquid uncured one. The addition of hardener must also contribute to the observed decrease in conductivity through promoting the formation of small isolated agglomerates of graphene, which were observed by ourselves and others [4]. After the oven heating step the conductivity increased again (~ $3 \times 10^{-4}$ S/m), which we associated with decreases in the viscosity of the systems when the materials are heated above the $T_g$, which could lead to thermally-induced relaxation of the graphene network. We define 'thermally-induced relaxation of the graphene network' here as an increase in the mobility of the GNP in the matrix happening above $T_g$, which is evidenced by DSC and will be further discussed in the *Thermal properties of the GNP/epoxy composites* section), which results in an increased number of interparticle contact points after 'reshaping' of the nanocarbon network as previously observed and reported for CNTs [14-16].

*3.2. Out-of-Autoclave (OoA) curing method based on Joule heating*

The formation of an electrically conductive GNPs network integrated in the epoxy matrix is essential for the development of a resistive heating OoA curing method. In our experiments, when an electric current was passed through a GNP/epoxy composite at loadings above the percolation threshold, a fast rise in temperature was observed through Joule heating of the graphene nanoparticles as they are close enough to each other. An infrared camera was employed to monitor the increase in the temperature of the material under the electrical field. The thermographs (shown in Fig. 3) revealed a heating rate of ~ 13 °C/min for the material under the applied electric field, whereas much lower heating rates (~ 3 - 4 ºC/min) are typically associated with the use of a standard oven. The intensity of the electric current applied through the composite was carefully controlled in order to keep the temperature constant at 100 ± 5 ºC for 4 h (as this was the length employed for the oven-based curing method). In addition to a rapid increase in temperature to reach 100 °C, Fig. 3 also reveals that by applying an electric field to the mixture a highly uniform temperature was achieved in the whole composite. Up to ~50 °C the temperature was very homogenous throughout the whole volume prepared, whereas at higher temperatures a slightly inferior uniformity was observed. Fig. 3 shows higher temperatures at the centre of the mould (right in the part where the current in applied) relative to more peripheric points. (All the studies in this work were performed on the central part of the cured material in order to have total certainty that the characterized material has been actually cured at the desired temperature).

*3.3. Thermal properties of the GNP/epoxy composites.*

DSC analysis was employed to evaluate the degree of curing of GNPs/epoxy composites at 10 wt.% loading of GNPs prepared by the oven-based and OoA curing methods and the results are shown in Fig. 4. No exothermic peak could be found in any case, which evidenced a complete cure of both composite materials. Passing an electric current through the composite material was thus proved as effective as the conventional oven based method to fully cure GNPs/epoxy composites at loadings above the percolation theshold.

The $T_g$, defined here as the midpoint of enthalpy change, was found to be 85 °C and 78 °C for GNP/epoxy-OoA and GNP/epoxy-Oven cured composites, respectively. Both values were considerably lower than the $T_g$ of the pure epoxy (~ 100 °C), which we attributed to thermally-induced aggregation of the GNPs happening when the composite materials were heated above their $T_g$ during curing. Heating the composite materials above their $T_g$ is thought to increase the mobility of the GNPs in the matrix, probably due to a decrease in the viscosity of the materials, which leads to the formation of small agglomerates and microvoids at the interface between GNPs and polymer, and thus to a decrease in the $T_g$ of the composite with respect to the polymer. A loading of GNP as high as 10 wt.% must be also hindering the formation of cross-links in the resin [17]. In order to probe this theory, DSC of pure epoxy and 10 wt.% loaded composite after the first curing step of 24 hours at room temperature were performed (shown in SI). The presence of quite prominent exothermic peaks in both curves was observed, which suggested that the materials were not fully cured at this stage. In addition, $T_g$ ~ 44 °C and 46 °C were found for neat epoxy and 10 wt.%-GNPs/epoxy composite, respectively, which strongly supports our theory about a relaxation of the GNPs network and some thermally-induced aggregation occurring when the composite is heated, which leads to a decreased $T_g$ value with respect to that found for oven cured epoxy. (It should also be noted that the $T_g$ found after the first curing step at room temperature were considerably lower than those found for the fully cured materials, as expected).

*3.4 Microstructure of the GNP/epoxy composites*

3.4.1 Dispersion of the GNPs in the matrix

Raman mapping analysis were performed on the GNPs/epoxy composites cured in the traditional oven in order to investigate the dispersion of the GNPs in the polymer matrix and the results obtained are shown in Fig. 5. Fig. 5 also shows the Raman spectra of the as-received GNPs, showing the features typical of graphene nanoplatelets, *i.e.* D band, G band and 2D band located at frequencies of 1360 cm$^{-1}$, 1590 cm$^{-1}$ and 2690 cm$^{-1}$, respectively. The low intensity of the D band with respect to the G band indicated the presence of small amount of defects. The intensity, position and shape of the 2D band was typical of graphene nanoplatelets with more than 10 layers.

The Raman spectrum of a typical oven cured GNP/epoxy composites showed a combination of features from the GNPs and from the epoxy resin, with the ratio of intensities between GNPs bands and the epoxy bands depending on the GNPs loading. The quality of the dispersion of the GNPs in the matrix was determined by evaluating the ratio of intensities between the G band of GNPs and the 1609 cm$^{-1}$ band of epoxy taken as the reference, i.e. $I_G/I_{1609}$. The Raman mapping results for GNP/epoxy composites with 0.3 wt.%, 5 wt.% and 10 wt.% loading of GNPs are shown in Fig. 5. A quantitative analysis from the mapping data was obtained by calculating the standard deviation (*D*) of the ratio of intensities studied here for each GNP loading, using the Equation:

$$D = \sqrt{\frac{\sum(r_i - r_a)^2}{N}} \qquad \text{Equation (3)}$$

where $r_i$ is the ratio of intensities obtained for each single spectrum taken, $r_a$ is the average value of the ratio of intensities and *N* is the total spectra taken for the mapping. The values obtained for *D* from Raman mapping performed at different loadings of GNPs were found to be 0.1703, 0.3176 and 0.1958 for 0.3 wt.%, 5 wt.% and 10 wt.%, respectively. The small values obtained for *D* revealed the presence of well distributed GNP in the polymer matrix at the loadings studied here. It is thought, however, that well distributed small agglomerates must be present in the high loaded composites and must contribute positively to the electrical properties of the oven-cured composites. Raman mapping analysis confirmed thus good dispersions of the GNPs in the matrix, with very small levels of aggregation even at high loadings.

SEM images of the fracture surface of pure epoxy and the GNP/epoxy composites prepared here are shown in Fig. 6S4. For the pure epoxy a very smooth morphology could be observed, whereas 'rougher' morphologies were found for the oven cured composites, with increasing 'roughnesses' observed for higher loadings. This clearly suggested that higher fracture energies were required to break the composite specimens relative to pure epoxy ones. From the SEM images the presence of GNPs could be easily distinguished from the polymer matrix due to their typical features, such as curved surfaces and folded or blended flakes. Although the density of GNPs observed on the fracture surfaces of the composites was found to increase with loading, good dispersions of the flakes (and/or well dispersed small agglomerates of them) in the matrix were found in all the composites, even at loadings as high as 10 wt.%.

A comparison between high-resolution SEM images of oven-cured and OoA-cured composites at 10 wt.% loading of GNPs (Figs. 6g,h) revealed that less micro-voids and a more compact structure were observed when the material was cured under an electric field. We attribute this induced changes in microstructure to the small separation between the graphene flakes at 10 wt.% loading, as a consequence of which the heat flow results in very fast heating rates and a considerable more uniform curing.

3.4.2 Orientation of the GNPs in the matrix

Li *et al.* [13] developed a method in which polarized Raman spectroscopy was employed to evaluate and quantify the orientation of graphene materials in bulk composites (described in the *Experimental Section*). This method was used here to analyze the orientation of the GNPs in the epoxy matrix for the oven cured and the OoA cured composites with 10 % loading of GNPs in *x* and *z* directions (Fig. 7). The GNPs were found randomly oriented in the oven cured sample, with the Raman G band ($I_G$) being independent of the rotation angle, in agreement with previous work [5].

However, for the OoA-cured composites, although the $I_G$ was independent of the angle of rotation in one direction ($\Phi z$), $I_G$ showed a strong dependency in the perpendicular direction ($\Phi x$), with maximum intensities found at 0º and 180º and minimum intensities at 90º. This result evidenced that the application of an electric field promoted an orientation of the GNPs in the matrix in the direction of the applied field, rendering an anisotropic composite. This result gives light to the possibility of tuning the microstructure of GNPs/epoxy composites to meet specific requirements for practical applications, including electrical and mechanical properties.

*3.5. Electrical properties of GNP/epoxy-OoA composites*

A comparison between the electrical behaviour of GNP/epoxy-Oven and GNP/epoxy-OoA composites at 10 wt.% loading is shown in Fig. 2. The oven cured samples are isotropic in their conductive nature, in agreement with the Raman results for the particle orientation. However, the OoA samples showed a degree of anisotropy with the conductivity one order of magnitude higher in the direction of the applied curing electric field, which again correlates with the polarized Raman studies. The preferential network formation in the direction of an electric field and the subsequent increase in conductivity has been observed previously by Wu *et al.* [18], who reported the use of an electric field to obtain an enhanced alignment of the GNPs in thermoset composites.

*3.6. Mechanical properties of the GNP/epoxy composites.*

The tensile testing curves of epoxy and oven cured composites at all the loadings studied revealed a glassy behavior, with the GNPs providing reinforcement of stiffness (Fig. S3). As shown in Fig. 8, a linear increase of the Young's modulus of the epoxy with GNPs content up to an optimal loading of ~ 4 wt.% was found, showing a maximum increment of 21 % with respect to pure epoxy. Above the optimal loading the modulus did not increase any further, due possibly to the formation of agglomerates of graphene in the matrix through van der Waals forces and π-π interactions, which deteriorated the mechanical properties, as previously reported for similar systems [19]. The rule of mixtures has been widely employed to successfully determine the effective modulus of graphene materials in similar systems [19, 20]. A modified rule of mixtures was applied here to determine the effective modulus ($E_f$) of the fillers in the nanocomposites using:

$$E_c = E_f V_f + E_m V_m \qquad \text{Equation (4)}$$

where $E_c$ is the Young's modulus of the composite, $E_f$, $E_m$ are the Young's modulus of the filler and matrix, respectively, and $V_f$ and $V_m$ are the volume fractions of the filler and the matrix, respectively, within the nanocomposite (with $V_f + V_m = 1$). The effective graphene modulus $E_{\text{eff}}$ could be calculated from $E_f$ using:

$$E_f = E_{\text{eff}} \eta_0 \eta_l \qquad \text{Equation (5)}$$

where $\eta_0$ is Krenchel orientation factor that depends on the average orientation of the particles [21] and $\eta_l$ is the length parameter which accounts for poor stress transfer at the filler-matrix interface for particles with small lateral dimensions ($\eta_l = 1$ for perfect stress transfer and $\eta_1 = 0$ for no stress transfer). For this calculation we employed $\eta_0 = 8/15$ [21] since polarized Raman revealed a random orientation of the GNPs in the matrix (*Section 3.3*), and $\eta_1 = 1$ assuming a perfect stress transfer. From the slopes of the lines up to the optimal loading in Fig. 6b values of $E_f = 14$ GPa and $E_{eff} = 26$ GPa were determined for the nanofiller used here. Although the tensile testing results revealed good levels of reinforcement for these GNPs epoxy composites, the modulus found for these nanoplatelets is much lower than their estimated 300 GPa value [22]. For the GNP/epoxy-Oven composites, the stress and strain at break (Fig. S3) were not found to increase with the addition of GNPs relative to pure epoxy in the range of loadings studied here, in agreement with previous works on similar systems, which is typically related to the presence of small agglomerates of graphene at the interface. It is well known that that the tensile strength in graphene/polymer composites is highly sensitive to the presence of small agglomerates and it has been discussed somewhere else [19]. The decrease in tensile strength observed here for the composites relative to pure epoxy must be related to the presence of small agglomerates of graphene in the polymer matrix. The relatively weak interface typically existing between non-functionalized fillers and polymers must be also contributing to the observed decrease in tensile strength. The values and standard deviation of the Young's modulus, tensile strength and tensile strain of all the studied composites are compiled in Table S1.

Given the need to be sufficiently above the percolation threshold (~ 8.5 wt.%) to use the OoA method, only a 10 wt.% GNP composite was prepared for mechanical testing by Joule heating. The stress-strain curves and obtained Young's modulus for oven and OoA cured

GNPs/epoxy composites at 10 wt.% loading are compared in Figs. 8a,b. A slightly higher Young's modulus (4.8 % increase) was found for the OoA composite relative to the oven cured one. In addition, although still below the pure epoxy values, considerably higher stress (48.5 %) and strain (60 %) at break were found for the OoA-cured composites. The improved mechanical properties observed for the OoA cured material relative to the traditionally oven cured one at identical loading must be related to the differences in microstructure found, which includes a more compact structure with less amount of microvoids and, more importantly, the orientation of the flakes induced in the matrix when the material is cured under an electric field (the current was applied along the length direction of the dumbbell specimens during curing and the tensile tests were indeed performed along the same direction, which is the direction of alignment). The latter will, of course, change the Krenchel orientation factor for the system.

The overall mechanical performance found for the GNPs/epoxy systems fabricated here is inferior to those observed for the carbon-fiber/epoxy composites typically employed in the aerospace sector, and they should not be seen as competitors to replace them. In contrast, this work probes that the Joule heating effect can be successfully employed to in-situ cure graphene/epoxy composites, which potentially could have great impact for novel emerging applications in different industrial sectors (including, but not exclusive to, the aerospace one) in which additional functionality (which only graphene materials can provide, such as barrier properties or outstanding EMI shielding performances) might be required, in addition to good mechanical and electrical properties.

Besides, the simplicity and low cost of this method could make it potentially attractive to industry. Although there are still many factors to take into account in the calculation of the total cost of these GNPs/epoxy nanocomposites (in addition to the energy cost, S5), the energy saving calculated already represents an important point to take into account and envision this method as potentially industrially viable. Some limitations related to the method still need to be overcome though. For example, further investigation on the mixing method to reduce the amount of solvent and sonication/shear mixing times, as well as on the manipulation of mixtures with high viscosity and other factors that at the moment might represent important limitations to scale it up, is still strongly required, though, in order to further reduce costs before this method can be industrially viable.

Similar resistance induced OoA cure protocols have been already successfully applied to fibre reinforced epoxy composites and pre-preg laminated composite materials typically used in aerospace [2]. The application of electrically induced cure methods to hybrid systems composed of carbon-fibres and GNPs could potentially hold an extraordinary interest as hierarchical materials with exceptional mechanical and electrical properties. The use of carbon fibers would considerably reduce the percolation threshold with respect to that one found here using only GNPs as fillers, whereas the GNPs would be expected to considerably enhance the adhesion between filler and host polymer. In addition, the use of graphene as additive could also facilitate the OoA manufacturing of carbon fibre composites. The development of novel OoA cure methods which are adapted to these and others hierarchical materials could open the door to new revolutionary applications in different industrial sectors.

## 4. Conclusions

A scalable Out-of-Autoclave method to cure structural thermoset composites through Joule heating of graphene networks integrated in the matrix was developed. The electrical behaviour of both uncured and oven cured GNPs/epoxy composites at loadings up to 10 wt.% were first evaluated, finding a percolation threshold of ~ 8.5 wt.% for the oven cured system. Electron microscopy and Raman spectroscopy revealed good dispersions of the GNPs in the matrix even at higher loadings. Consequently, above the percolation threshold the well-dispersed electrically conductive GNPs networks were found to act as nanoheaters integrated in the matrix through simple Joule heating when an electric current was passed through them. DSC revealed that the Joule heating led to similar degrees of curing to those obtained by the traditional oven heating.

The electrically-induced heating resulted in much faster heating rates and more uniform curing than that typically associated for oven heating, which led to more compact structures with less microvoids. In addition, polarized Raman revealed a preferred orientation of the GNPs in the direction of the applied current, which was further evidenced by enhanced electrical and mechanical properties in the direction of alignment. The microstructure induced in the OoA cured composites led to superior electrical and mechanical properties relative to those found for oven cured materials. The Joule heating based OoA in-situ curing method developed here for GNP/epoxy composites emerges thus as highly promising to fabricate structural mechanically reinforced and electrically conductive graphene based thermoset composites for novel applications such as in the aerospace sector.

**Acknowledgements.** The authors are grateful to the European Union Seventh Framework Programme (grant no: 604391 Graphene Flagship) for financial support.

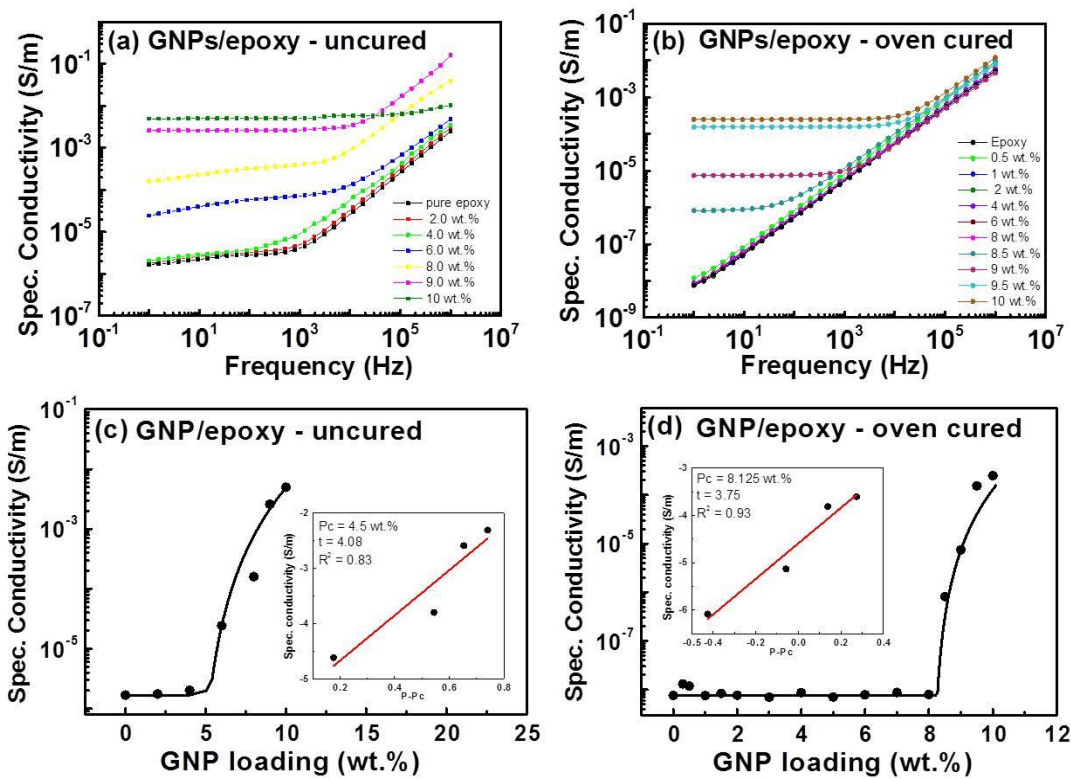

**Fig. 1.** Log-log plot of conductivity as a function of frequency for uncured (a) and oven cured (b) samples. Semi-log plot of conductivity as a function of GNP loading for uncured (c) and oven cured (d) samples. (The inserts are log-log plots of conductivity as a function of $P-P_c$).

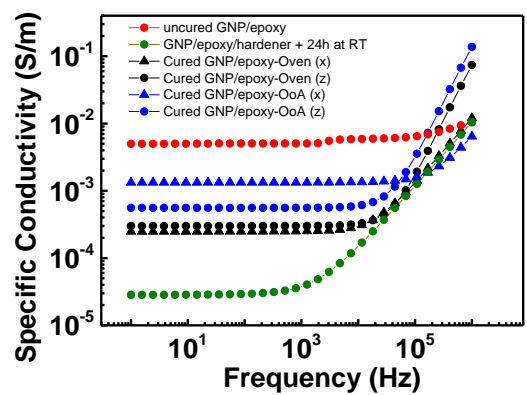

**Fig. 2.** Log-log plot of specific conductivity as a function of frequency for GNP/epoxy-Oven and GNP/epoxy-OoA cured composites (at 10 wt.% loading of GNPs) at different stages of curing. The conductivity along two perpendicular directions ($x$, $z$) was tested for both oven- and OoA-cured composites.

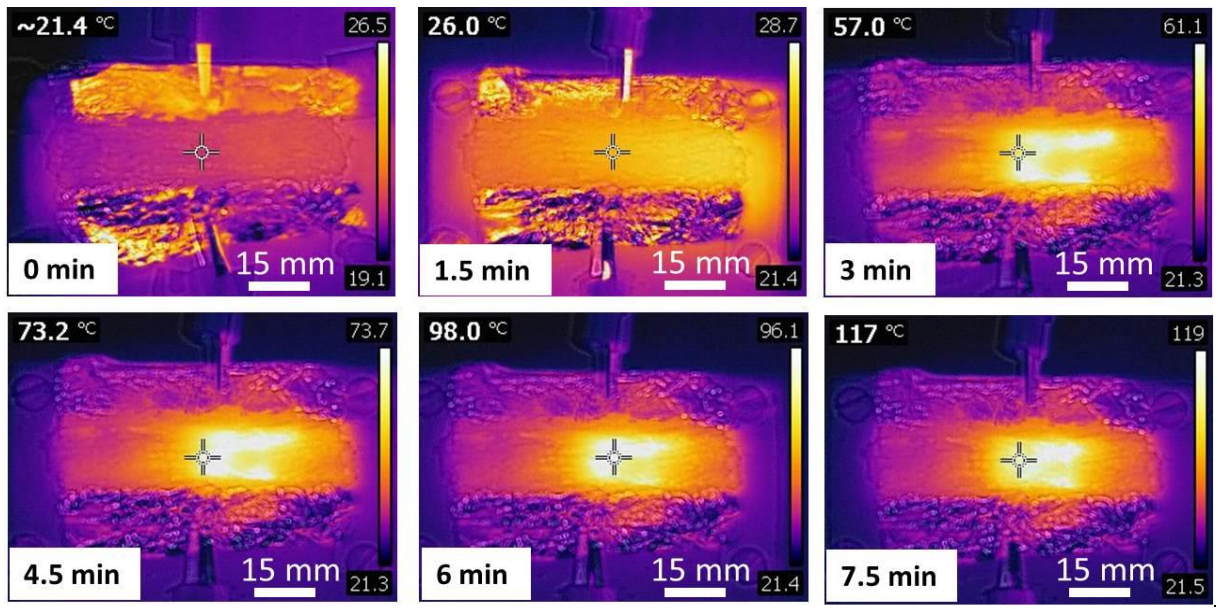

**Fig. 3.** Thermographs showing the variation of the temperature of the GNP/epoxy composite with time when curing under an electric field.

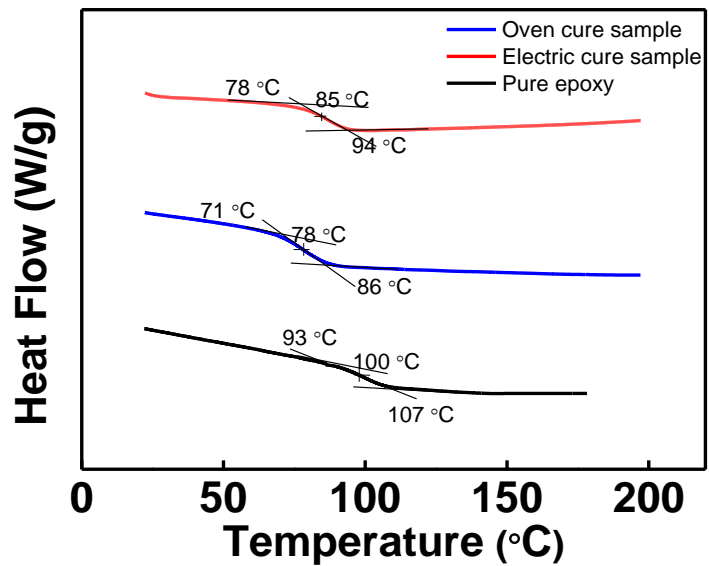

**Fig. 4.** DSC results for Oven and Out-of-Autoclave cured epoxy composites with 10 wt.% loading of GNPs.

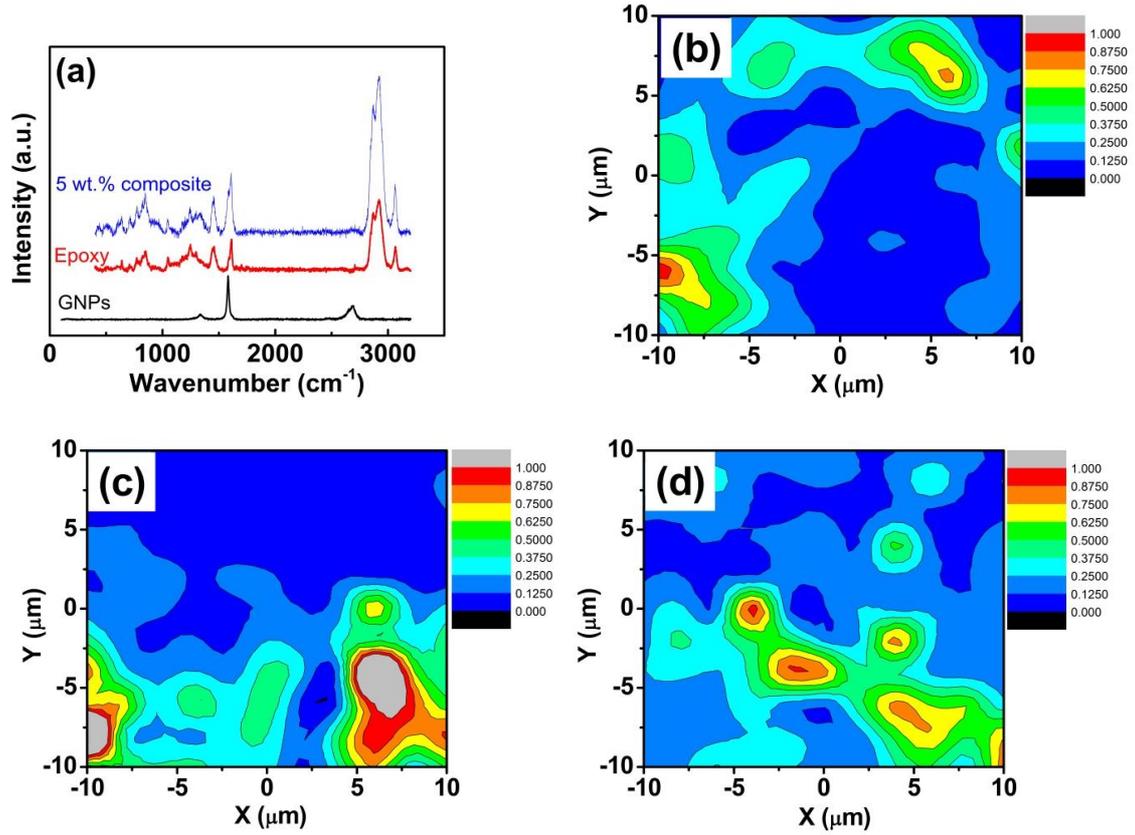

**Fig. 5.** Raman spectra of as-received GNPs, pure epoxy and an oven cured GNP/epoxy composite (a). Raman mapping of oven cured GNP/epoxy composites with 0.3 wt.% (b), 5.0 wt.% (c) and 10.0 wt.% (d) loading, showing the variation of the values found for $I_{G(GNPs)}/I_{1609 cm^{-1}(epoxy)}$ in 20 μm × 20 μm areas of the composites.

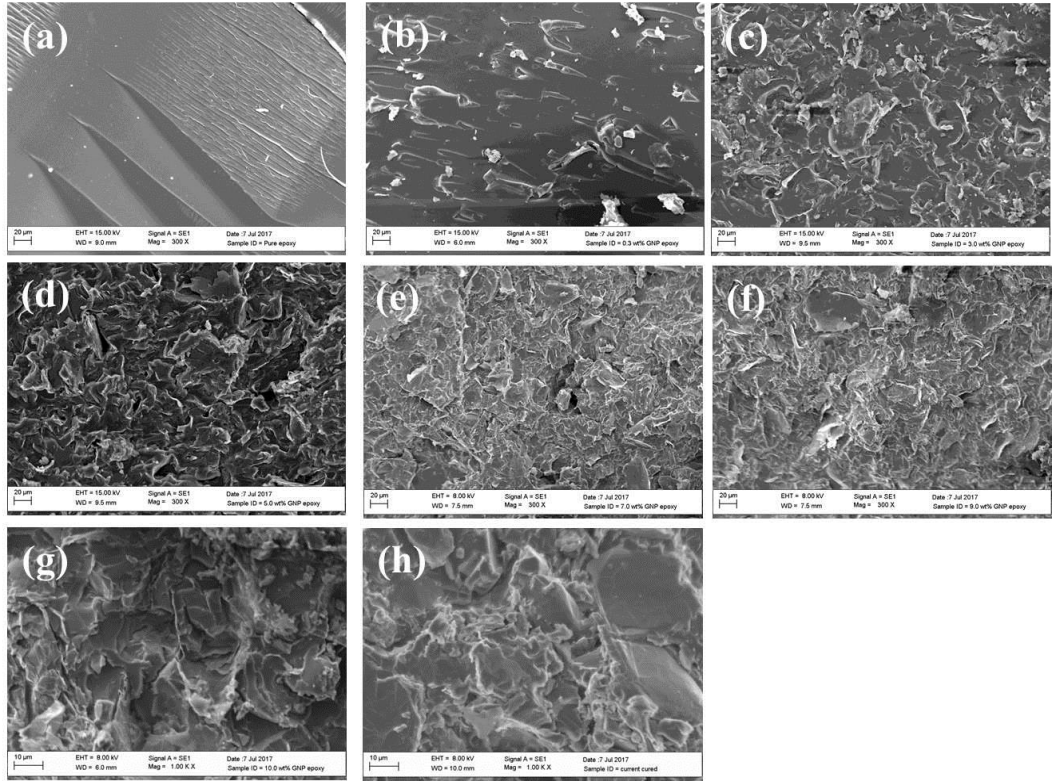

**Fig. 6.** SEM images of the fracture surface of the pure epoxy (a), GNP/epoxy-Oven composites with 0.3 wt.% (b), 3 wt.% (c), 5 wt.% (d), 7 wt.% (e), 9 wt.% (f), 10 wt.% (g) loading of GNPs; GNP/epoxy-OoA composite at 10 wt.% loading of GNPs (h).

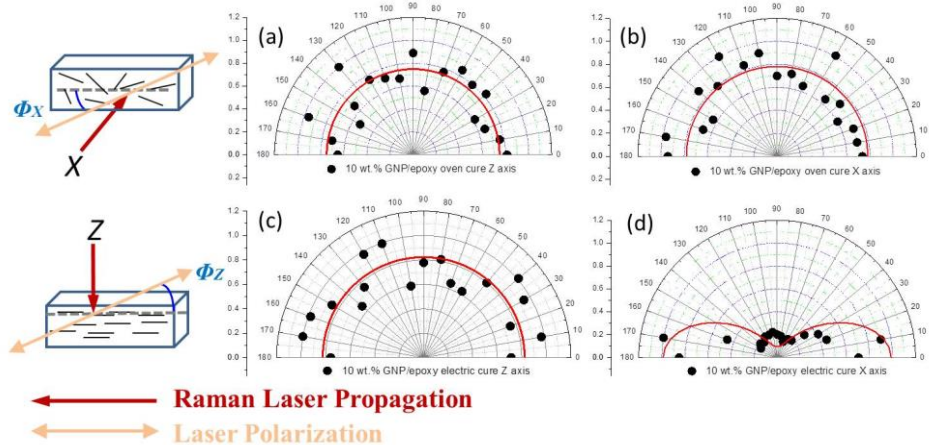

**Fig. 7.** Variation of $I_G$ as a function of $\Phi z$ (a) and $\Phi x$ (b) for 10 wt.%-GNP/epoxy-Oven composite. Variation of $I_G$ as a function of $\Phi z$ (c) and $\Phi x$ (d) for 10 wt.%-GNPs/epoxy-OoA composite. Schematic description of the *x*, *y* and *z* directions in the specimens.

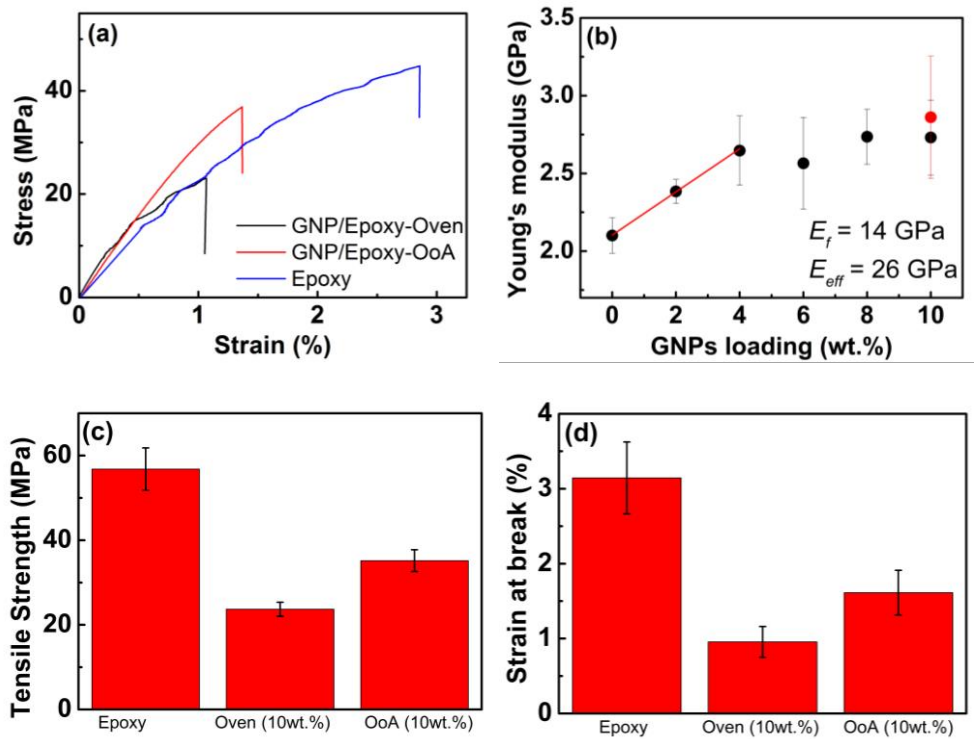

**Fig. 8.** (a) Stress-strain curves for epoxy, GNPs/epoxy-Oven and GNPs/Epoxy-OoA composites (10 wt.%). (b) Variation of the Young's modulus with loading for GNPs/epoxy-Oven and comparison with the GNP/Epoxy-OoA (red data point at 10 wt.%). Stress (c) and strain (d) at break of epoxy, GNPs/Epoxy-Oven and GNP/Epoxy-OoA composites.